\input harvmac.tex


\def\HH{{\cal H}}
\def\HHtilde{{\widetilde\HH}}

\Title{NSF-KITP-07-107}
{\vbox{\centerline{Minimal Length Uncertainty Relations and}
\centerline{New Shape
Invariant Models}}}

\centerline{Donald
Spector\footnote{$^\dagger$}{spector@hws.edu} }
\medskip\centerline{Department of Physics, Eaton Hall}
\centerline{Hobart and William Smith Colleges}
\centerline{Geneva, NY \ 14456 USA}

\vskip .3in
This paper identifies a new class of shape invariant models.  These models are
based on extensions of conventional quantum mechanics that
satisfy a string-motivated minimal length uncertainty relation.  An important
feature of our construction is the pairing of operators that are not adjoints
of each other.  The results
in this paper thus show the broader applicability of shape invariance to exactly
solvable systems.
\vskip .5in
\centerline{PACS: 03.65.Fd, 02.30.Hq, 11.30.Pb }

\Date{3/2007}

\newsec{Introduction}

Shape invariance has proven to be a useful paradigm for understanding exact
solvability in quantum mechanics \ref\shapeinv{L. E. Gendenshtein, {\it JETP Lett.} 38 (1983) 356.}
\ref\shaperev{F. Cooper, A. Khare, and U. Sukhatme, {\it Phys. Rep.} 251
(1995) 267.}.  It
is of interest to understand how broadly shape invariance can be applied, and in
this paper, I will identify new arenas in which it can be used effectively.

In the past few years, there has been some consideration of extensions of quantum 
mechanics designed to comport well with
string 
theory \ref\takeuchi{L.N. Chang, D. Minic, N. Okamura, and T. Takeuchi, 
{\it Phys. Rev.} D65 (2002) 125027.}.  In particular, string theory predicts that
there can be a modification in the
familiar Heisenberg uncertainty 
relationship \ref\newUR{D. J. Gross and P. F. Mende, {\it Nucl. Phys.} B303 (1988) 407\semi
D. J. Gross and P. F. Mende, {\it Phys. Lett.} B197 (1987) 129\semi
A. Kempf, G. Mangano, and R. B. Mann, {\it Phys. Rev.} D52 (1995) 1108 (1995).}, so that 
it takes the form
\eqn\modunc{
\Delta x \ge {a\over \Delta p} + b \Delta p~~~.}
This relationship is sometimes termed a ``minimal length uncertainty
relationship,'' since it implies a lower bound for $\Delta x$, namely
$\Delta x \ge \sqrt{a/b}$.
Of course, in order for such a relation to appear in the non-relativistic
limit, the Schr\"odinger equation must be modified.  It is precisely such
modifications that are considered in \takeuchi , in which some exactly
solvable models are found that
satisfy the minimal length uncertain relationship \modunc .  

It is the point of this letter to show that the exact solvability of these models
is no accident.  These models turn out to be representatives of a new class of
shape invariant models.  In this letter, I will identify these new shape
invariant models, and use the shape invariance to obtain solutions.  One will see
that these models include the models considered by \takeuchi, as well as some
spectral problems that generalize the Schr\"odinger equation in ways beyond
the modification \modunc.  An explicit
consequence of this result is 
to re-affirm the close connection between shape invariance and exact
solvability for spectral problems in general, and not simply in the
context of the Schr\"odinger equation.  We also make technical progress,
demonstrating that shape invariance can be applied productively in a
setting in which second-order differential
operators can be written as the product of pairs of first-order differential operators that 
are not adjoints of each other.

\newsec{One-dimensional generalized harmonic oscillator}

We begin by displaying the simplest model of \takeuchi . To begin, \takeuchi\ considers a modified commutation relation
\eqn\modcomm{[\hat x, \hat p] = i\hbar (1+\beta \hat p^2)~~~.}
Such a commutation relation leads to the uncertainty relationship
\eqn\commbetah{\Delta x \ge {\hbar\over 2} \Bigl({1\over \Delta p}+ \beta \Delta p\Bigr)~~~.}
This commutation relation can be realized in the momentum representation by
operators
\eqn\opform{\eqalign{\hat x &= i\hbar\left[ (1+\beta p^2){d\over dp}\right]+\gamma p\cr
\hat p &= p}}
We note that the parameter $\gamma$ is a trivial addition here; its presence does not
serve to modify the conventional commutation relation, and it is easily set to zero
by a canonical transformation.  It is  included here not simply for making contact with
\takeuchi, but more importantly to lay the groundwork for one of the considerations
needed for the multi-dimensional oscillator considered in the next section.

Quantizing the classical one-dimensional harmonic oscillator
\eqn\classicalonedimHO{H ={1\over 2\mu}p^2+ {1\over 2}\mu\omega^2 x^2}
according to \modcomm\ produces the momentum space Schr\"odinger-like 
time-independent equation
\eqn\takho{
\Bigl[ -\mu\hbar\omega\Bigl\{\Bigl( (1+\beta p^2){\partial\over\partial p}\Bigr)^2
 +2\gamma p\Bigl( (1+\beta p^2){\partial\over\partial p}\bigr) 
 +\gamma(\beta+\gamma)p^2 + \gamma\Bigr\} + {1\over\mu\hbar\omega}p^2\Bigr]
 \Psi(p) = {2E\over \hbar\omega}\Psi(p)~~~.}
 Remember, too, that the physics cannot depend on $\gamma$.
 
 We now wish to demonstrate that this equation can be solved exactly via shape
 invariance techniques.  This provides a simpler understanding of the exact solvability
 (and a simpler calculation of the energy levels and wavefunctions)
 demonstrated by analytical means in \takeuchi .
  
A quick review of shape invariance is in order here; more detailed treatments are found in the
literature, including a review \shaperev\ and a recent article 
revealing the underlying algebraic structure of shape invariance as a BPS phenomenon
associated with centrally extended supersymmetry  \ref\shapeBPS{M. Faux and D. Spector, 
{\it J. Phys.} A37 (2004) 10397.}.  

Suppose one has a Hamiltonian $H = {\bar A} A$ for which the spectrum is bounded from below by
zero.  In the familiar applications, $\bar A = A^\dagger$, which enforces the positive 
semi-definiteness of the spectrum, but as long as the spectrum is positive semi-definite, $\bar A$
need not be the adjoint of $A$.   The eigenfunctions for such Hamiltonians with eigenvalue zero are those annihilated by $A$.  The partner Hamiltonian $\widetilde H = A {\bar A}$ will have the same
spectrum of non-zero eigenvalues as $H$, since if $H\phi = \Lambda \phi$, then
$\widetilde H (A\phi) = \Lambda (A\phi)$.  Typically, $A\phi=0$ has solutions and
$\bar A\bar\phi = 0$ does not, and the spectral degeneracy does not extend to the
zero eigenvalues.

Shape invariance exists when $H$ and $\widetilde H$ have the same mathematical form, so
that $\widetilde H(g) = H(g^\prime)+c$, where $g$ denotes the coupling constants of the theory,
$g^\prime$ are the transformed coupling constants which are functions of the $g$, and $c$
is some $c$-number, which also generally depends on the 
parameters of the theory.  This means that the energies of the two Hamiltonians not only
exhibit the degeneracy referred to in the preceding paragraph ($E_{k+1} = \widetilde E_k$), but
also satisfy the relation $\widetilde E_k(g) = E_k(g^\prime)+c$.  These results enable us to find
the energy eigenvalues algebraically.

The wavefunctions are also found algebraically, through standard raising and lowering operator
techniques.   The ground state of $H$ is the solution of $A\phi = 0$; the states $\widetilde \phi_k$ of $\widetilde H$
are related to the states $\phi_k$ of $H$ by $\widetilde\phi_k(g) = \phi_k(g^\prime)$ as well
as by $\phi_{k+1} = {\bar A}\widetilde\phi_k$, which allows the eigenstates to be built up algebraically
from the ground state of $H$.

We now demonstrate the solution of \takho\ via shape invariance.  To begin, define
the operators
\eqn\AAdefine{\eqalign{A &=  (a+bp^2){d\over dp} + cp \cr
   {\bar A} &= -(a+bp^2){d\over dp} + cp~~~.}}
The real numbers $a$, $b$, and $c$ are all positive.  Note that with this 
convention, there is a normalizable state annihilated by $A$ but not one
annihilated by ${\bar A}$.

Consider the equation $\HH\phi = \Lambda \phi$ where $\HH={\bar A} A$.
Then
\eqn\firstH{\HH(a,b,c)= -\left[(a+bp^2){d\over dp}\right]^2 +c(c-b)p^2- ac ~~~.}
Similarly, $\HHtilde = A {\bar A}$ is
\eqn\secondH{\HHtilde(a,b,c) = -\left[(a+bp^2){d\over dp}\right]^2+c(c+b)p^2+ca~~~.}
Consequently, $\HH$ and $\HHtilde$ form a shape invariant pair, since
\eqn\shapefirst{\HHtilde(a,b,c) = \HH(a,b,c+b)+2ca+ba}
Applying the principles of shape invariance, one readily obtains the eigenvalues for
$\HH(a,b,c)\phi_k(p)=\Lambda_k\phi_k(p)$ 
and $\HHtilde(a,b,c)\phi_k(p)=\widetilde\Lambda_k\widetilde\phi_k(p)$.
One finds $\Lambda_0=0$, $\widetilde\Lambda_0 = a(b+2c) = \Lambda_1$, 
$\widetilde\Lambda_1 = a(b+2(c+b))+a(b+2c) = \Lambda_2$, and so forth, resulting 
in the general expression
\eqn\eigenvaluesHO{\Lambda_k = a(k^2b+2kc)\qquad\qquad k=0,1,2,3,\ldots~~~.}

We can easily make contact with previous results.  Note that, as pointed out above, the
parameter $\gamma$ has no physical content, and thus may readily be set to zero.
Once this is done, the modified harmonic oscillator
equation \takho\ of \takeuchi\ is equivalent to $m\HH\phi(p) = \Lambda\phi(p)$, with
the identifications
\eqn\takhoID{\eqalign{m&=\mu\hbar\omega\qquad\qquad a=1\qquad\qquad b=\beta\cr
c(c-b)&={1\over (\mu\hbar\omega)^2}
\qquad\qquad\qquad \Lambda + mca = {2E\over \hbar\omega}}}
This leads to the energy eigenvalues for \takho
\eqn\agreeonedim{E_k=
\hbar\omega\left[ {1\over 2}\mu\hbar\omega\beta (k^2+k+{1\over 2})
   + (k+{1\over 2})\sqrt{1+{\beta^2\mu^2\hbar^2\omega^2\over 4} }~\right]}
which agrees with the past results.  It is a simple exercise, too, to obtain the
wavefunctions.

Note that if we wish to use shape invariance to solve \takho when the term linear in ${d\over dp}$ 
is explicitly present, it is still possible to
do so, as we now demonstrate.

Suppose we define modified raising and lowering operators
\eqn\notadjAAdag{\eqalign{
A &= F(p){d\over dp} + W(p)+\Omega(p) \cr
{\bar A} &= -F(p){d\over dp} +W(p) - \Omega(p)~~~.}
}
Then
\eqn\notadjHHtilde{\eqalign{
\HH &= {\bar A} A = -\left[F(p){d\over dp}\right]^2
-F(p)\left({dW\over dp}+{d\Omega\over dp}\right)-2F(p)\Omega(p){d\over dp}+W^2(p)-\Omega^2(p)\cr
\HHtilde &= A {\bar A} = -\left[F(p){d\over dp}\right]^2
+F(p)\left({dW\over dp}-{d\Omega\over dp}\right)-2F(p)\Omega(p){d\over dp}+W^2(p)-\Omega^2(p)~~~.}}
To make contact with the problem at hand, we set
\eqn\tellFWOmega{
F(p) = (a+bp^2)\qquad\qquad W(p) = c_1 p \qquad\qquad \Omega(p)=c_2 p~~~,
}
in which case
\eqn\nowHHis{
\HH = -\left((a+bp^2){d\over dp}\right)^2 -2(a+bp^2)c_2 p {d\over dp}
   +(c_1^2-c_1b-c_2^2-c_2b)p^2-c_1a-c_2a~~~,}
which now includes the ${d\over dp}$ term.
Computing $\HHtilde = A{\bar A}$, we find 
\eqn\nowshapeis{
\HHtilde(a,b,c_1,c_2) = \HH(a,b,c_1+b,c_2)+2ac+ba~~~.}
This result is identical in form to the shape invariance result \shapefirst;
the addition of $\Omega=c_2 p$ thus has no bearing on the eigenvalue
spectrum.  Even in considering the wavefunctions, there is no physical modification associated with 
including this term proportional to $\gamma$, simply a change of variables.

\newsec{Multi-dimensional Oscillator}

We now generalize our results to another shape invariant eigenvalue equation.  This will turn
out to be the relevant equation for the ``radial'' (in momentum space) piece of the generalized 
Schr\"odinger equation that describes the $D$-dimensional harmonic oscillator with a minimal length uncertainty relation, the second of the two models discussed in \takeuchi .

To obtain our new shape invariant model, we define
\eqn\multiAA{
\eqalign{A &= (a+bp^2){d\over dp} + cp + {g\over p}\cr
{\bar A} &= -(a+bp^2){d\over dp} + cp + {g\over p}~~~.}
}
Then we can compare $\HH = {\bar A} A$ and $\HHtilde = A {\bar A}$.  One finds
\eqn\multiHH{
\eqalign{
\HH(a,b,c,g) = -\left[(a+bp^2){d\over dp}\right]^2 + {g(g+a)\over p^2} + c(c-b)p^2 +c(g-a)+g(c+b)\cr
\HHtilde(a,b,cg) = 
-\left[(a+bp^2){d\over dp}\right]^2 + {g(g-a)\over p^2} + c(c+b)p^2 +c(g+a)+g(c-b)
}
~~~.
}
The shape invariance is obvious, with
\eqn\multiID{
\HHtilde(a,b,c,g) = \HH(a,b,c+b,g-a) + 4a(b+c) - 4 bg~~~.
}
The eigenvalues for the equation $\HH\phi_k(p) = \Lambda_k\phi_k(p)$
are readily obtained algebraically.  They are given by
\eqn\multieigen{\Lambda_k = 4abk^2 + 4(ac-bg)k~~~.}

As it stands, this equation is not quite the momentum-space radial equation for the
modified $D$-dimensional harmonic oscillator.  That equation, given in
\takeuchi, is
\eqn\theirlongequation{
\eqalign{
-\mu\hbar\omega\left[\left\{[1+(\beta+\beta^\prime)p^2]{d\over dp}\right\}^2
+\left\{{D-1\over p}+[(D-1)\beta + 2\gamma]p\right\}
\left\{[1+(\beta+\beta^\prime)p^2]{d\over dp}\right\} \right.\cr
\left.- {L^2\over p^2}
+(\gamma D - 2 \beta L^2)
+\{\gamma(\beta D+\beta^\prime+\gamma)-\beta^2L^2\}p^2\right] R(p)
+{1\over \mu\hbar\omega}p^2R(p) = {2E\over \hbar\omega}R(p)}}
which is obtained from quantizing the classical $D$-dimensional harmonic
oscillator Hamiltonian 
$H = {1\over 2}\mu\omega^2\vec x\cdot\vec x+{1\over 2\mu}\vec p\cdot\vec p$ 
according to the modified commutation relation
\eqn\multicomm{
[\hat x_i,\hat p_j] = i\hbar (\delta_{ij}+\beta{\hat p}^2\delta_{ij}+\beta^\prime{\hat p}_i{\hat p}_j)~~~,}
which is done using the representations
\eqn\multixandp{\eqalign{
{\hat x}_i &= i\hbar\left[(1+\beta p^2){\partial\over \partial p_i}+\beta^\prime p_ip_j
{\partial\over \partial p_j}+\gamma p_i\right]\cr
{\hat p}_i &= p_i
}
}
Note that, as in the previous example, there is no physical content to $\gamma$ and,
it may simply be set to any value.

The problem, however, in comparing \multiHH with \theirlongequation, is that $\HH$ 
in the model we have written down has no term linear in ${d\over dp}$, yet such a
term cannot be removed from the multi-dimensional oscillator equation simply by
re-defining $\gamma$, as was the case in one dimension.  
Fortunately, the same method used
to include the term linear in ${d\over dp}$ in the previous section can be used productively here.

We once again define modified $A$ and ${\bar A}$ operators, namely
\eqn\modifiedAAdag{\eqalign{
A&= (a+bp^2){d\over dp}+(c_1+c_2)p+{g_1+g_2\over p}\cr
{\bar A} &= -(a+bp^2){d\over dp} +(c_1-c_2)p + {g_1-g_2\over p}
}
}
Defining, as usual, $\HH={\bar A} A$ and $\HHtilde=A{\bar A}$, we obtain
\eqn\theHHthings{\eqalign{
\HH =
-\left((a+bp^2){d\over dp}\right)^2-(a+bp^2)(c_1-{g_1\over p^2})-(a+bp^2)(c_2-{g_2\over p^2})
 \cr-2(a+bp^2)(c_2p+{g_2\over p}){d\over dp}+(c_1p+{g_1\over p})^2 - (c_2p+{g_2\over p})^2\cr
\HHtilde =
 -\left((a+bp^2){d\over dp}\right)^2+(a+bp^2)(c_1-{g_1\over p^2})-(a+bp^2)(c_2-{g_2\over p^2})
\cr -2(a+bp^2)(c_2p+{g_2\over p}){d\over dp}+(c_1p+{g_1\over p})^2 - (c_2p+{g_2\over p})^2
 }
 }
If $c_2=g_2=0$, then we simply recover the operators of \multiHH; however,
the terms with $c_2$ and
$g_2$ in \theHHthings\ are the same in both $\HH$ and $\HHtilde$, and thus 
there is a 
shape invariance relationship between $\HH$ and $\HHtilde$ in this new scenario
which is identical to the
relationship that followed from \theHHthings, even when $c_2$ and $g_2$ are non-zero.  Note, though, that with $c_2$ and
$g_2$ non-zero, $\HH$ and $\HHtilde$ have exactly the same mathematical form
as \theirlongequation, the equation for the minimal uncertainty relationship $D$-dimensional harmonic oscillator.  Consequently, we can use the shape invariance of the new $\HH$ and 
$\HHtilde$ to obtain the solutions to \theirlongequation.  In particular, we have
\eqn\longshape{\eqalign{
\HHtilde(a,b,c_1,c_2,g_1,g_2) &= \HH(a,b,c_1+b,c_2,g_1-a,g_2)+4a(b+c_1)-4bg_1 \cr
\Lambda_k &= 4abk^2 + 4(ac-bg)k~~~.}
}
These $\Lambda_k$ give the spectrum for the modified $D$-dimensional oscillator, with the
appropriate identification of parameters.  Likewise, the ground state is found by setting
$A\phi=0$, and the excited states by the application of ${\bar A}$.

\newsec{Additional Shape Invariant Models in this class}

In this section, I will describe the extension of the shape invariance techniques
developed here to models that fit into the present analysis, but that go beyond
that considerations of \takeuchi\ and the simple minimal length
uncertainty principle.

Suppose we have an eigenvalue equation $H\phi = E\phi$ where $H$ is given by
\eqn\extendedsinh{
H = -\left[(a+b\sinh^2y){d\over dy}\right]^2 + {\gamma\over \cosh^2y}~~~,
}
and the parameters $a$, $b$, and $\gamma$ are all positive.
Here, I have deliberately labeled the variable $y$ to emphasize that the technique
described herein is not restricted to a particular physical framework.
Let us define
\eqn\sinhAAdag{\eqalign{A&=(a+b\sinh^2y){d\over dy} + g \tanh y \cr
{\bar A}&=-(a+b\sinh^2y){d\over dy} + g \tanh y~~~.}}
Then
\eqn\sinhHH{\HH(a,b,g) = {\bar A} A
=-\left[(a+b\sinh^2y){d\over dy}\right]^2 -{g(g+a-b)\over \cosh^2y}+g(g-b)~~~,
}
where $g$, like $a$ and $b$, is positive.
The partner operator $\HHtilde = A{\bar A}$ is related to $\HH$ by
\eqn\sinhshape{
\HHtilde(a,b,g) = \HH(-a,-b,g) = \HH(a,b,g+b-a) + (2g-a)(a-b)~~~.
}
Thus this model exhibits shape invariance under the transformation
$g\rightarrow g+b-a$.  Using the standard techniques, we see that
the eigenvalues $E_k$ for the operator $\HH$ are simply
\eqn\sinhEvalues{
E_k = k(2g-b)(b-a)-k^2(a-b)^2\qquad\qquad k=0,1,2,\ldots
}
Note that for large enough $k$, these energies become negative, which we know
is not physically allowed.  This is not special to the example at hand.  In fact, if
we consider ordinary quantum mechanics with potential $1/\cosh^2(x)$, there is 
only a finite number of bound states; these are the states extracted by the standard
shape invariance argument, 
and when the energy goes negative, it is the signal that we have left the realm of
physical states.  There is always at least one normalizable state, which is the state
with vanishing energy in the supersymmetric/shape invariance framework.
This entire structure persists in the problem at hand.

Note, too, that it is simple to make contact with the original operator $H$ 
in \extendedsinh\ above.
To be precise,
\eqn\HHHrelation{
H(a,b,\gamma) = \HH(a,b,\bar g)-\bar g(\bar g-b)
}
where $\bar g = {1\over2}[(b-a)+\sqrt{(b-a)^2+4\gamma}~]$.

Finally, we note that if generalize this procedure to equations given by
$\HH = {\bar B} B$ where
\eqn\boringgen{B = \Bigl(1+f(y)\Bigr){d\over dy}+g(y)
\qquad
{\bar B} = -\Bigl(1+f(y)\Bigr){d\over dy}+g(y)~~~,
}
then we can always identify models in which 
\eqn\aretheseboring{
f(y) ={\alpha g^2(y) + \beta\over g^\prime(y)}
}
for which $B\bar B$ has the same mathematical form as ${\bar B} B$,
and the spectra are positive semi-definite, so that the shape invariance
approach is applicable.

\newsec{Conclusions}

In this paper, I have demonstrated that shape invariance can be applied effectively to
exactly solvable models in extensions of quantum mechanics.  These include models
that possess a modified uncertainty relationship, namely the minimal length uncertainty relation, which also emerges in string theory.  As usual, shape invariance provides a simpler and clearer understanding of the exact solvability of the models in question.  We have found, too, that shape invariance
can be effectively applied even for spectral problems in which the differential operator
being studied must be written as the product of lower order differential operators that are
not adjoints of each other.  

We note, too, that given the connection between centrally extended supersymmetry and shape invariance demonstrated in \shapeBPS , the present work implies a supersymmetric construction that can incorporate these minimal-length uncertainty principle extensions of quantum mechanics, and the 
use of BPS techniques to extract the states in these theories.  The exploration of those issues,
however, lies beyond the scope of this paper.

\newsec{Acknowledgments}
This research was supported in part by the National Science Foundation 
under Grants No. PHY04-57048 and PHY05-51164, by the Japan Society for the Promotion
of Science.  Portions of this work were conducted at RIKEN (The Institute for Chemical and Physical Research) and at the Kavli
Institute for Theoretical Physics, both of which I thank for their
hospitality.

\listrefs
\bye